# Scattering Suppression of Silica Microspheres with Semicontinuous Plasmonic Shell

Vashista C. de Silva[1], Piotr Nyga[2][**], and Vladimir P. Drachev[1][*]

[1]*University of North Texas, Department of Physics and Center for Advanced Research and Technology*
*1155 Union Circle, Denton, TX, USA 76203*
[2]*Institute of Optoelectronics, Military University of Technology*
*2 Kaliskiego Str., 00-908 Warsaw, Poland*

The scattering and absorption suppression of the silica microspheres is experimentally achieved by semicontinuous gold shell synthesized on the silica microspheres. These fractal gold shells on the silica spheres are shown to have significant extinction in the visible and mid-infrared spectral range. It is found that the silica's scattering peak in the visible at 560 nm and the relative contribution of the vibrational stretching band at 9 μm range gradually decreases with increasing of the gold coverage.

Light scattering by core-shell particles made of dielectric and metal results in a variety of phenomena predicted many decades ago [1–3]. By varying the relative dimensions of the core and continuous shell, the optical resonance of these nanoparticles can be varied over hundreds of nanometers in wavelength, across the visible and into the infrared region of the spectrum [4,5]. Fractal shells below the percolation threshold exhibit geometrically tunable plasmon resonances due to the silver nanoparticle arrangement [6]. With shell growth, the plasmon resonance can be tuned from a single dipolar mode in sparse coatings to multiple high order shell excitations in densely packed thick shells [6]. Invisibility for coated confocal ellipsoids was introduced by Kerker [2] in the approximation of long wavelength planar electromagnetic waves. The corrections could be made proportional to the eighth power of the wavenumber by adjusting both the permittivity and permeability of the shell and core of the coated spheres [3]. It was found in [7] that suitably fabricated core-shell spheres do not have to be too small compared to the incident wavelength to achieve sensible invisibility for planar electromagnetic waves; the scattering cross-section could be relatively small for core-shell spheres with cores with sizes comparable to one fifth of the wavelength based on the scattering-cancellation properties of plasmonic materials with low positive or negative permittivity [8–14]. Moderately sized dielectric or conducting objects may be effectively cloaked with a simple homogeneous and isotropic layer of plasmonic material. In this case, the wave can penetrate through the shell and the effect is associated with the out-of-phase scattering properties of plasmonic materials. This helps to cancel out the core's positive polarizability with shell's negative polarizability hence creating scattering cancellation of silica microspheres. These concepts then extended to collections of particles [11], multi-frequency operation [12], and larger sized objects [14]. The negative polarizability of the plasmonic shell can be resulted from multiple scattering centers due to their random arrangement [15]. Another way to minimize radiation from an obstacle can be realized by choosing a boundary control function on a part of its boundary [16]. It was also found - that a certain fish scale structure of copper strips can be invisible for polarized normal incident plane waves at given frequency [17]. In contrast with the coordinate-transformation methods [18–21], as pointed out in [22,23] the cloaking can be achieved in the region external to the cloak, exploiting the metamaterial resonances between the cloak and the background. The fields outside the cloaking region and inside the core are not perturbed by the presence of the source, due to the anomalous localized resonances induced in the shell regions and in the cloak [22,23]. First experimental realization of plasmonic cloaking based on a scattering-cancellation technique due to the local negative polarizability of metamaterials was reported at microwave frequencies [24,25]. An array of metallic fins embedded in a high-permittivity fluid has been used to create a metamaterial plasmonic shell capable of cloaking a dielectric cylinder.

In this paper we study scattering of silica microspheres coated with a gold semicontinuous plasmonic shell in the broad optical range, from 400 nm to 20 μm, covering shorter and longer wavelengths relative to the microsphere size. Comprehensive spectral measurements of sub-monolayers of core-shell particles on planar substrates show scattering suppression in the visible range accompanied by the increase in total transmission. The results indicate that this suppression is not just a spectral shift of the resonance in scattering, but suppression of its amplitude without noticeable shift. In the infrared range the gold semicontinuous shell "hides" the absorption resonance of the silica sphere at 9 μm. Performed here simulations with an effective [26,27] continuous shell do not provide a full correspondence to the experiments, while qualitatively reflect important features. This emphasizes the key role of semicontinuous nature of the shells in the observed spectra. Note that such type of semicontinuous geometry has fractal dimension [28-33].

The optical properties of the metal-dielectric semicontinuous films are influenced by surface plasmon resonances (SPRs) [27] in metal nanostructures, accumulating and building up electromagnetic energy in a broad spectral range at nanometer scale [28–33]. As it is known the critical value of the metal coverage, called percolation threshold [28] results in a variety of plasmon resonances covering a spectral range from the visible to infrared. The main idea here behind the presented experiments is to utilize broad band response and efficient coupling of radiation of semicontinuous shell at percolation threshold to explore the possibility of scattering cancelation for core-shell spheres with dimensions comparable to the wavelength of incident light. In this paper we show the experimental realization of scattering and absorption suppression of silica microspheres by coating them with fractal gold nanostructures.

In the experiments silica microspheres with diameter of about 800 nm were coated with gold nanostructure using a modified method of the reduction of gold salt $HAuCl_4$ with formaldehyde in the presence of surfactants and stabilizers, initially developed for continuous shells [34]. In order to fabricate semicontinuous shells the seeding of silica surface with gold nanoparticles step was skipped and gold was directly synthesized on the silica microspheres. The surface coverage of the gold on the microsphere surface was varied for different samples by varying the reduction time. Specifically, 30 mg of $K_2CO_3$ has been dissolved in 100mL of ultra-pure water, the solution was stirred for 10 minutes, and 300µL of 50 mM $HAuCl_4$ was added in the solution. Next, $8 \times 10^8$ of the amine functionalized silica microspheres and then 40µL of 0.36 mM formaldehyde were added to the vigorously stirred solution. After the growth of the gold shells, solution was stabilized with polyvinylpyrrolidone (PVP) to prevent aggregation of the microspheres. Finally solution was centrifuged to collect gold-coated silica microspheres. Collected particles were re-dispersed a few times in ultra-pure water in order to remove contaminants. The synthesized gold-coated microspheres were deposited on zinc selenide (ZnSe) substrates for the infrared spectroscopy and on fused silica substrates for spectroscopy in the visible range. Silica microspheres were purchased from Bangs Laboratories, Inc. All other chemicals, PVP, chloroauric acid ($HAuCl_4$), $K_2CO_3$, and formaldehyde were purchased from Sigma-Aldrich.

The structural characterization was performed with Hitachi S-4800 field emission scanning electron microscope (FESEM). To collect the optical spectra in 0.32µm-2.5µm and 2.5µm-20µm wavelength ranges, we used Perkin Elmer Lambda 950 spectrometer and Nicolet Fourier-transform infrared (FTIR) spectrometer respectively. The reflectance, and scattering measurements were performed with the integrating sphere module of the Perkin Elmer Lambda 950 spectrometer [35].

Several different samples have been synthesized to study the effect of the shell structures with gradually increasing gold coverage. Figure 1 shows FESEM images of the samples under study. The data for normalized extinction (NE) are presented in Fig. 1a and 1b. The normalized transmittance (NT) spectra were first normalized by the transmittance of the bare substrate and then the normalized extinction (NE) spectra were calculated as NE= (-log (NT))/$N_S$, where $N_S$ is the particle surface density calculated from the FESEM images and taken in µm$^{-2}$. FESEM images of different shell structures and corresponding NE spectra are presented in Figure 1. The samples are labeled from 0 to 4, where 0 stands for the sample of bare silica microspheres ($N_S$=0.4±0.015) and 1-4 correspond to the samples with gradually increasing gold coverage in the shells ($N_{S2}$=0.66±0.02, $N_{S3}$=0.57±0.02, $N_{S4}$=0.5±0.02). For the bare silica microspheres there is the absorption peak due to the Si-O-Si vibrational stretching band at 9µm [36], and there is an extinction peak in the visible spectral range due to Mie scattering resonances.

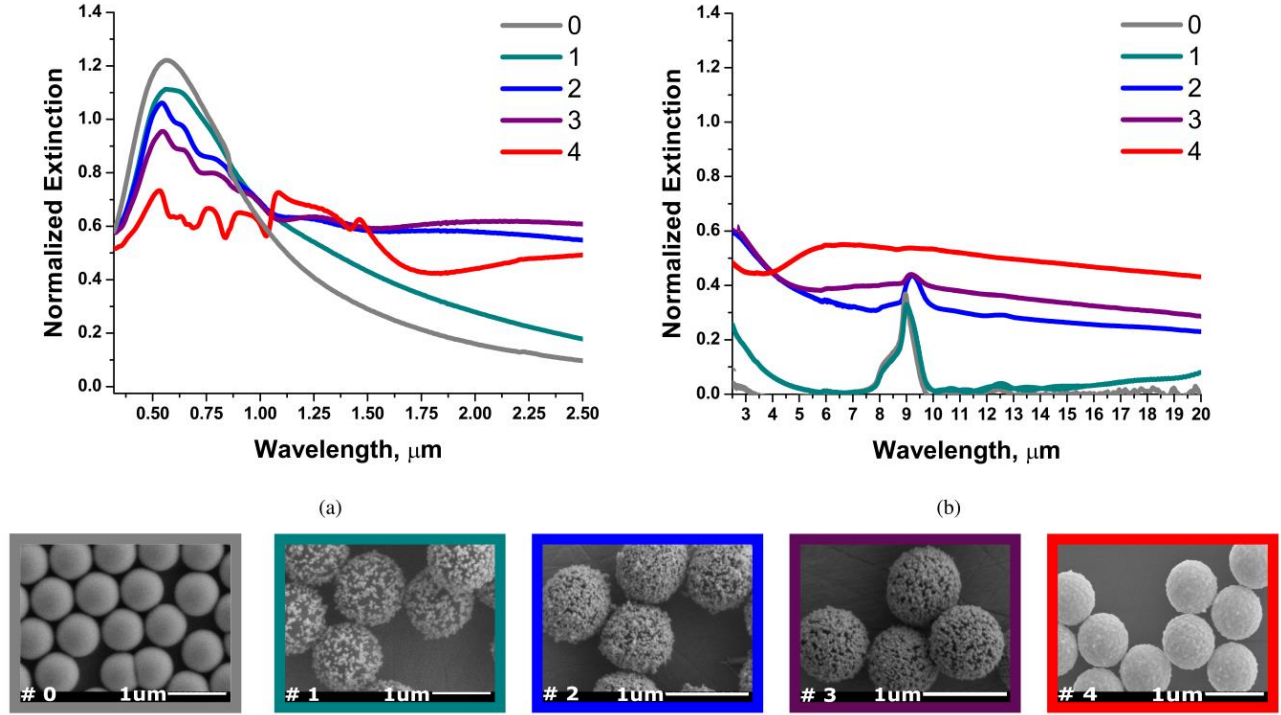

(Color online) Figure 1: Normalized extinction spectra and FESEM images of the core-shell particles with different gold shell coverage and morphologies.

The spectra "1"-"4" show strong dependence on the shell structure. The extinction spectra become broadband and spread out for the visible to infrared as the coverage increases. Note that this feature looks similar to the planar semicontinuous films, where the broadening of extinction spectra has been theoretically and experimentally demonstrated at the percolation threshold [30,31,33]. The shell structure of the sample "1" is composed mainly of small Au fractals and most of them are isolated. That is the reason why the extinction does not increase much for longer wavelengths as this shell structure is well below the percolation threshold. As soon as the percolation threshold is achieved, as for the sample "2", the extinction increases for the infrared region relative to the shell structure "1". The extinction increases for both bellow and above Si-O-Si vibration band for shell structure "2". By keeping the shell thickness about the same and increasing the metal filling fraction as shown in "3", the extinction is increased at the wavelength of Si-O-Si vibration band. The extinction of $SiO_2$ sphere at the position of the Si-O-Si vibration band is caused mainly by the absorption in silica and scattering does not have a major effect. Note that we have observed the broadband extinction for different shell thicknesses. In figure 1, shell structure "4", the shell thickness is 38 nm compare to 25 nm in "3", "2" and "1" shell structures. The shell structure "4" is above the percolation threshold and shows higher extinction than "3" due to increase in shell thickness and metal filling fraction.

Note that silica microspheres have low absorption in the visible, the extinction from bare silica particles is purely due to the scattering. The sample "1", with the shell made of isolated small aggregates of gold particles, shows decrease in extinction as compared to sample "0". As the gold coverage of the shell structure increases creating gold fractals, the extinction decreases further.

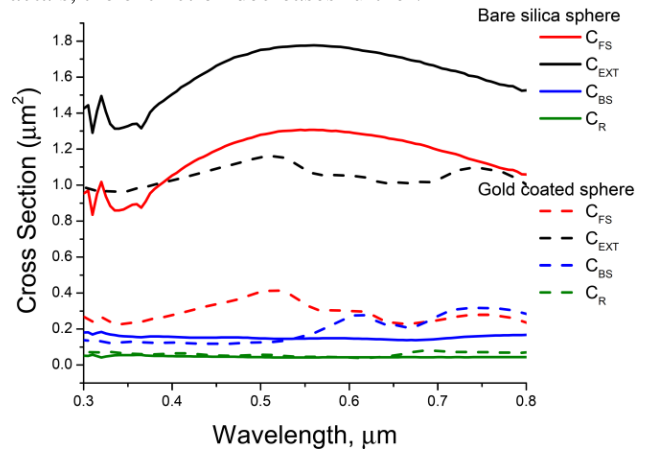

(Color online) Figure 2: Measured cross sections, $C_{FS}$, $C_{EXT}$, $C_{BS}$, and $C_R$ spectra of bare silica microsphere and gold-coated silica microsphere "4".

One can see that the broadening occurs for the samples 2-4. Surprisingly it completely cancels the absorption at silica vibration band at 9 μm. The scattering peak at 560 nm is suppressed to the level of fractal shell absorption in the visible range.

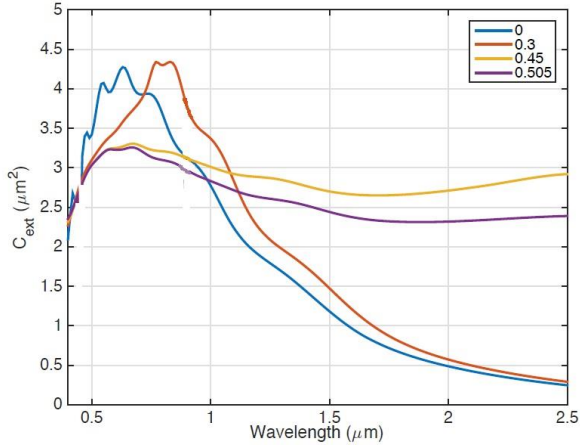

(Color online) Figure 3. Extinction cross section spectra for different shell gold coverage (0, 0.3, 0.45, 0.505) simulated for the visible-near IR spectral range.

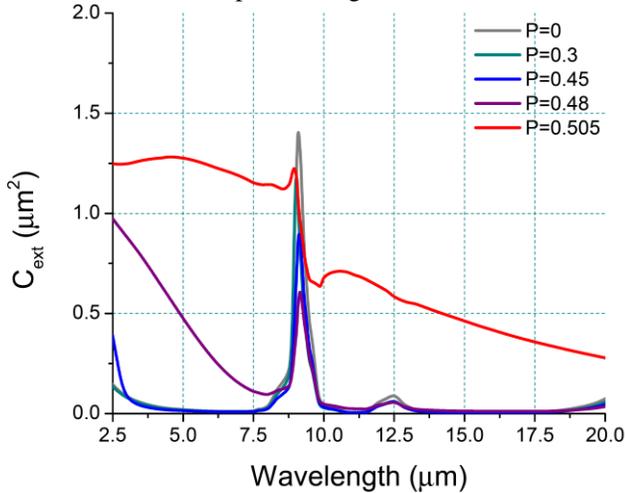

(Color online) Figure 4. Extinction cross section spectra for different shell gold coverage simulated (0, 0.3, 0.45, 0.48, 0.505) for the IR spectral range.

In order to "itemize" the source of extinction, the transmittance, reflectance, and forward- and back-scattering spectra were collected for bare silica and gold-coated silica microspheres, i.e. sample "0" versus sample "4". The spectra were collected with the integrating sphere accessory [35]. We measured the transmittance (T) along the optical axis, the forward scattering (FS), which is the scattering in the forward semi-sphere excluding transmitted along the axis, the reflectance (R) - the light reflected at small angle, backward scattering (BS), which is the scattering in the backward semi-sphere excluding the specular reflection. All the spectra are normalized per particle surface density, 0.4 μm$^{-2}$ for bare miscrospheres and 0.5 μm$^{-2}$ for core-shell sample "4". The results are plotted in Figure 2 for cross sections of the corresponding processes. For a single layer microspheres $\Delta I=-I_0 C_{ext} N_s$, where $I_0$ is the input light intensity, $C_{ext}$ is the extinction cross section and $N_s$ is the particle number density. Thus $C_{ext} \cdot N_s = 1-T$, $C_{FS}=FS/N_s$, $C_{BS}=BS/N_s$, and $C_R=R/N_s$, where $C_{ext}$, $C_{FS}$, $C_{BS}$, and $C_R$ are the extinction, forward-scattering, back-scattering, and reflection cross sections. The extinction cross section (black dashed line in Fig. 2) of the gold-coated sphere has decreased as compared to the bare silica microsphere (black solid line) near the Mie resonances. Also the forward scattering of gold-coated sphere (red dashed line) is decreased in average by about 75% compared to the bare silica microsphere (red solid line). The reflectance and back scattering have increased somewhat for the gold-coated microspheres compared to the bare silica microspheres.

In order to realize effect of scattering cancelation in the effective medium approximation the numerical simulations have been performed and presented in Figure 3 for the visible range and in Figure 4 for the infrared range.

To calculate the extinction cross section $C_{ext}$ of gold coated silica core-shell particles we used the Mie theory described in [37]. We model the particle as concentric system with silica core with radius $r_1$ and gold shell with thickness $t_1$ and radius $r_2=r_1+t_1$. The silica dielectric constant was taken from [38]. The gold shell in form of semicontinuous film with different coverage $p$ (from 0 to 1, where 0 corresponds to the case without shell and 1 to solid gold film) was calculated using Bruggeman effective medium approximation [26] using Drude model for gold dielectric constant [39]. The extinction cross section is given by:

$$C_{ext} = \frac{2\pi}{k^2}\sum_{n=1}^{\infty}(2n+1)\operatorname{Re}(a_n+b_n),$$

where $k=2\pi/\lambda$ is the wave vector, $\lambda$ is the wavelength in the ambient medium, and $a_n$ and $b_n$ are the scattering coefficients of electrical field. $a_n$ and $b_n$ were calculated using algorithm described in [40]. The sum runs for $n$ from 1 to ∞, but the series can be truncated at a some maximum $n_{max}$ ($n_{max} \approx k\,r_2$) [7,37]. To closer resemble the experimental conditions in IR spectral range (ZnSe substrate) we used n=2.4 for the refractive index of surrounding medium.

The numerical simulations show that the effective medium model works relatively well for long wavelengths. The model qualitatively reproduces the cancelation of the silica absorption at 9 μm by the gold shell with gradually increasing gold coverage. The model shows also scattering cancelation at the Mie resonance wavelength. However, the value of the extinction cancelation in the simulations is about 25% while in the experiments the extinction is suppressed for about 45% and the scattering is reduced by

75%. Obviously, the inhomogeneous shell has to be analyzed with more advanced models taking into account semicontinuous nature of the gold structure.

The scattering and absorption suppression of the silica microspheres by fractal gold nanostructures formed on the silica microspheres are demonstrated in this experiment.

To conclude, our experiments demonstrate scattering suppression of silica microspheres using plasmonic semicontinuous gold shells. Increasing the gold coverage can gradually decrease the scattering peak of silica microsphere in the subwavelength visible range. In the visible spectra, largest extinction was observed for bare silica microspheres due to Mie scattering. By measuring transmittance, reflectance, and forward- and back- scattering we found that transmittance is increased along with the scattering suppression for gold coated silica sphere. The scattering suppression with semicontinuous shell is more efficient relative to the predicted with the effective medium model for gold shell and core shell simulations. Also in the mid infrared spectra, it is shown that by using plasmonic fractal shell around silica microsphere, one can hide the Si-O-Si vibrational stretching band. As the gold coverage increases on the silica microspheres the relative contribution of the vibrational stretching band of silica at 9 μm in the total extinction is also decreased for the gold-coated silica microspheres.

Acknowledgments
This work was partially supported by The Edgewood Chemical Biological Center under the auspices of the U.S. Army Research Office.

Author to whom correspondence should be addressed.
*vladimir.drachev@unt.edu
** pnyga@wat.edu.pl.